\documentclass[a4paper, fleqn]{scrartcl}

\usepackage{amsmath}
\usepackage{amssymb}
\usepackage{graphicx}
\usepackage{setspace}
\usepackage[sort&compress]{natbib}

\usepackage[T1]{fontenc}
\usepackage{mathptmx}
\usepackage[scaled=0.90]{helvet}
\usepackage{courier}

\usepackage[sort&compress]{natbib}
  \bibpunct{[}{]}{,}{n}{}{;}
\usepackage[amssymb, mediumspace]{SIunits}

\usepackage{color}
  \definecolor{myDarkRed}{rgb}{0.5,0,0}       
  \definecolor{myDarkBlue}{rgb}{0,0,0.6}      
  \definecolor{myDarkGreen}{rgb}{0,0.4,0}     
  \definecolor{myDarkGray}{rgb}{0.4,0.4,0.4}  
  \definecolor{black}{rgb}{0,0,0}
  \definecolor{white}{rgb}{1,1,1}
  \definecolor{myRed}{rgb}{0.9,0.1,0.1}

\usepackage[%
  colorlinks={true},%
  linkcolor={myDarkBlue},%
  citecolor={myDarkGreen},%
  urlcolor={myDarkBlue},%
  pdfauthor={Isabel Krebs},%
  pdflang={en-US},%
  baseurl={},%
  breaklinks={true},%
  hyperfootnotes={false}%
  ]{hyperref}

\usepackage{doi}

\usepackage{lineno}
\linenumbers

\newcommand{\changes}[1]{{#1}}

\begin{document}

\setlength{\parindent}{0em}
\setlength{\parskip}{4ex plus2ex minus1ex}

\newcommand{\pderiv}[2]{\frac{\partial #1}{\partial #2}} 
\newcommand{\dderiv}[2]{\frac{d #1}{d #2}}               

\begin{center}
\textbf{\huge Nonlinear excitation of low-n harmonics in reduced magnetohydrodynamic simulations of edge-localized modes}

{\Large I.~Krebs$^1$, M.~H\"olzl$^1$, K.~Lackner$^1$, S.~G\"unter$^1$}

\textit{
$^1$Max Planck Institute for Plasma Physics, EURATOM Association, Boltzmannstr. 2, 85748 Garching, Germany 
}
\end{center}

\section*{Abstract}
Nonlinear simulations of the early ELM phase based on a typical type-I ELMy ASDEX Upgrade discharge have been carried out using the reduced MHD code JOREK. The analysis is focused on the evolution of the toroidal Fourier spectrum. It is found that during the nonlinear evolution, linearly subdominant low-n Fourier components, in particular the $n=1$, grow to energies comparable with linearly dominant harmonics. A simple model is developed, based on the idea that energy is transferred among the toroidal harmonics via second order nonlinear interaction. The simple model reproduces and explains very well the early nonlinear evolution of the toroidal spectrum in the JOREK simulations. Furthermore, it is shown for the $n=1$ harmonic, that its spatial structure changes significantly during the transition from linear to nonlinearly driven growth. The rigidly growing structure of the linearly barely unstable $n=1$ reaches far into the plasma core. In contrast, the nonlinearly driven $n=1$ has a rigidly growing structure localized at the plasma edge, where the dominant toroidal harmonics driving the $n=1$ are maximal and in phase. The presented quadratic coupling model might explain the recent experimental observation of strong low-n components in magnetic measurements [Wenninger et al., \emph{Non-linear magnetic perturbations during edge localized modes in TCV dominated by low n mode components}, submitted to Nuclear Fusion].


\newpage

\section{Introduction}\label{:intro}
Edge-localized modes~(ELMs) are relaxation-oscillation instabilities observed at the edge of tokamak plasmas in high-confinement regime~(H-mode). Ejecting energy and particles from the plasma, ELMs have the favorable effect of reducing the impurity content of the plasma and providing a mean to control the plasma density \citep{Zohm1996}. But if too large, they cause large heat fluxes which can damage plasma facing components \citep{Loarte2003,Klimov2011}. As the ability of controlling the ELM properties decides on whether the H-mode can be a suitable operational regime for ITER and future fusion reactors, the understanding of this instability is crucial. Nonlinear MHD simulations are an important tool in the quest for theoretical comprehension of ELMs. 

A nonlinear reduced MHD code which has been developed especially for edge instabilities, is the JOREK code \citep{Huysmans2007}. In this work, it is used for simulations of the early ELM phase, which are based on the geometry and parameters of an ASDEX Upgrade tokamak \citep{Hermann2003} discharge. Section \ref{:jorek} introduces JOREK and gives details about the simulations. The toroidal Fourier spectrum of the instability and its nonlinear evolution is analyzed and compared to recent experimental findings in Section \ref{:nonlinear}. It is observed, that initially weakly unstable toroidal Fourier components can become important nonlinearly.  In Section \ref{:model}, the question is addressed, what determines the nonlinear evolution of the toroidal harmonics in the simulations. A simple model is presented that shows how this evolution can be understood in the framework of second order nonlinear coupling between the toroidal harmonics. Finally, in Section \ref{:n1} it is investigated how the radial and poloidal localization of a linearly subdominant toroidal harmonic changes due to its nonlinearly driven growth. A summary and an outlook are given in Section \ref{:summary}.

\section{The JOREK code and the simulations}\label{:jorek}
\subsection*{JOREK}
The finite element code JOREK solves the nonlinear reduced MHD equations in full toroidal X-point geometry including separatrix and open flux surfaces. JOREK has originally been developed by G.T.A. Huysmans \citep{Huysmans2008,Czarny2008}. For the presented simulations, a single fluid version of JOREK\footnote{The used code revision is R706.} ("model302") has been used. The code is discretized via a Fourier decomposition in toroidal direction and 2D bi-cubic B\'ezier finite elements in the poloidal plane. The grid in the poloidal plane is aligned to the flux surfaces and can be refined in the regions of interest. The toroidal Fourier decomposition allows to choose the toroidal harmonics included in the computation. The discretization in time is performed according to a fully implicit Crank-Nicholson scheme. For the part of the boundary which follows the outermost open flux surface, ideally conducting wall boundary conditions are implemented, and for the divertor where the boundary is crossed by magnetic field lines, modified Bohm boundary conditions apply. The code uses a particular normalization of the physical variables. A JOREK time unit corresponds to approximately $0.5\,\mu$s at the parameters of the presented simulations. In the following, quantities indexed with "JOREK" are normalized according to the JOREK normalization scheme (units of these quantities are omitted). For the equations solved by the applied JOREK model and details about the normalization of the variables, please refer to Reference \citep{Hoelzl2012}.

\subsection*{The simulations}
The simulations are focused on the early ELM phase when the instability grows exponentially before the onset of nonlinear saturation. Emphasis has been put on the analysis of the nonlinear interaction of the toroidal Fourier harmonics, thus a large number of included toroidal harmonics and a high flexibility in combining them was required. The simulations are based on the simulations presented in Reference \citep{Hoelzl2012}\footnote{The same equilibrium and parameters as in the "eta5" simulation in Reference \citep{Hoelzl2012} have been used here.} with an additional modification of the code providing the possibility of excluding desired harmonics from the simulation. The baseline simulation includes 16 toroidal Fourier harmonics $n=1,2,...,16$ in addition to the axisymmetric $n=0$ part. To analyze the interaction of the different harmonics in more detail, a large number of simulations including different subsets of these harmonics has been carried out.

\begin{figure}
\centering
  \includegraphics[width=0.6\textwidth]{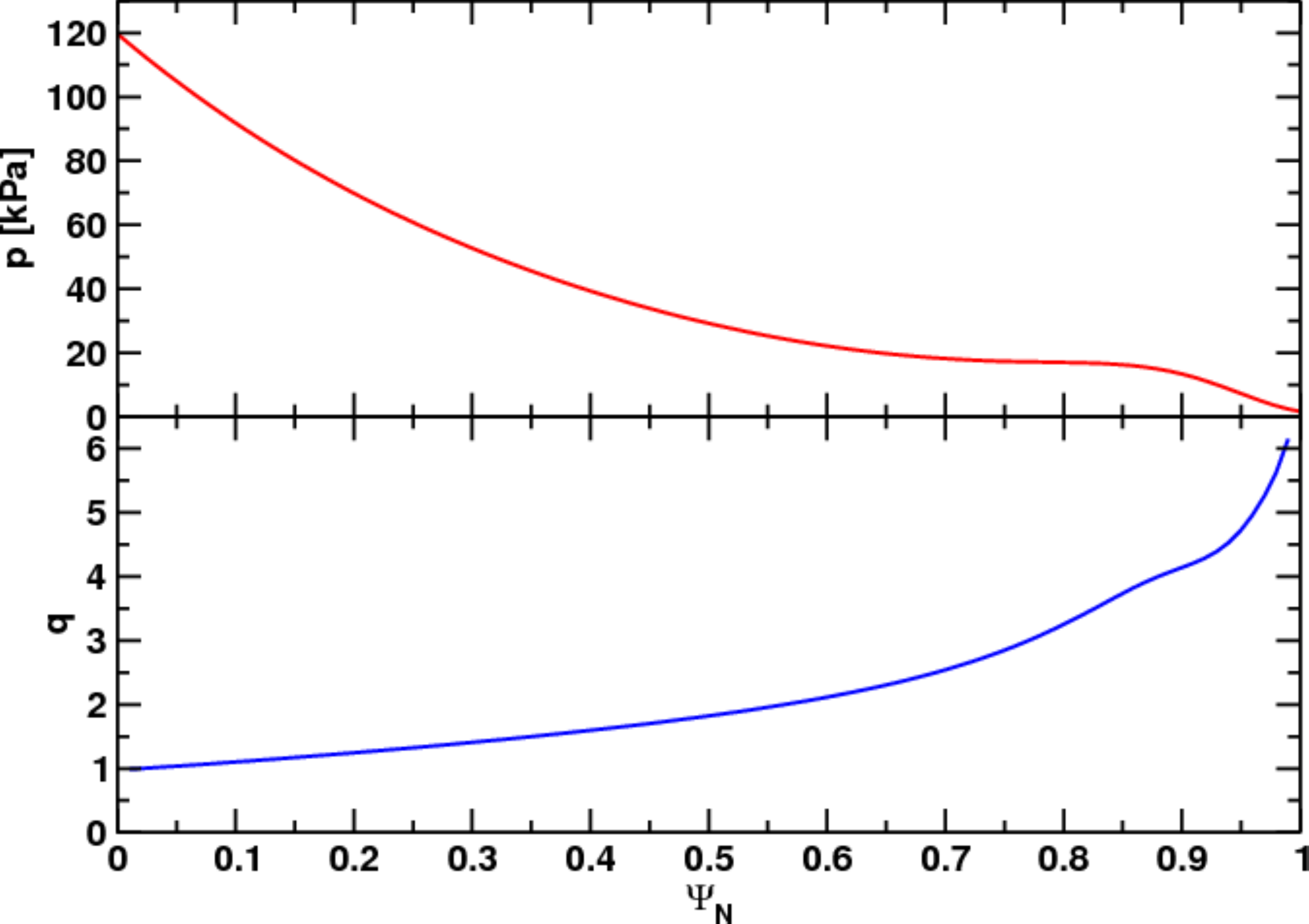}
\caption{Equilibrium pressure (red) and safety factor (blue), obtained from an equilibrium reconstruction of an ASDEX Upgrade discharge. The values of the safety factor in the center and at the edge are $q(0) \approx 1$ and $q(0.95) \approx 4.7$.}
\label{fig:profiles}
\end{figure}

The simulations are based on an equilibrium reconstruction of a typical type-I ELMy H-mode ASDEX Upgrade discharge (\#23221 at $4.7\,$s). The equilibrium reconstruction has been performed with the CLISTE code \citep{McCarthy1999,McCarthy1999b}. The corresponding equilibrium pressure and safety factor profiles are shown in Figure \ref{fig:profiles}. The particle density in the plasma center is $6 \cdot 10^{19}\, \textnormal{m}^{-3}$. Heat and particle sources and perpendicular heat and particle diffusivities are chosen such that the background profiles do not change significantly during the simulation. The parallel particle diffusivity is set to zero, parallel particle transport is thus provided by convection only. The heat diffusion anisotropy at the separatrix is about $\kappa_{\parallel}/\kappa_{\perp} = 7 \cdot 10^{6}$. Viscosity and resistivity have a $T_{N}^{-3/2}$ temperature dependency where $T_N$ is the temperature normalized by its value at the plasma center. The core viscosity is set to about $1.2 \cdot 10^{-5} \textnormal{kg} \textnormal{m}^{-1} \textnormal{s}^{-1}$. The resistivity ($\eta \approx 5 \cdot 10^{-5}\, \Omega$m in the core, leading to a Lundquist number of about $10^5$) is larger than in a realistic ASDEX Upgrade discharge\footnote{The core resistivity in ASDEX Upgrade discharges has values of about $10^{-8}\, \Omega$m.} due to computational restrictions.

\section{Nonlinear evolution of the toroidal harmonics}\label{:nonlinear}
\begin{figure}
\centering
  \includegraphics[width=0.6\textwidth]{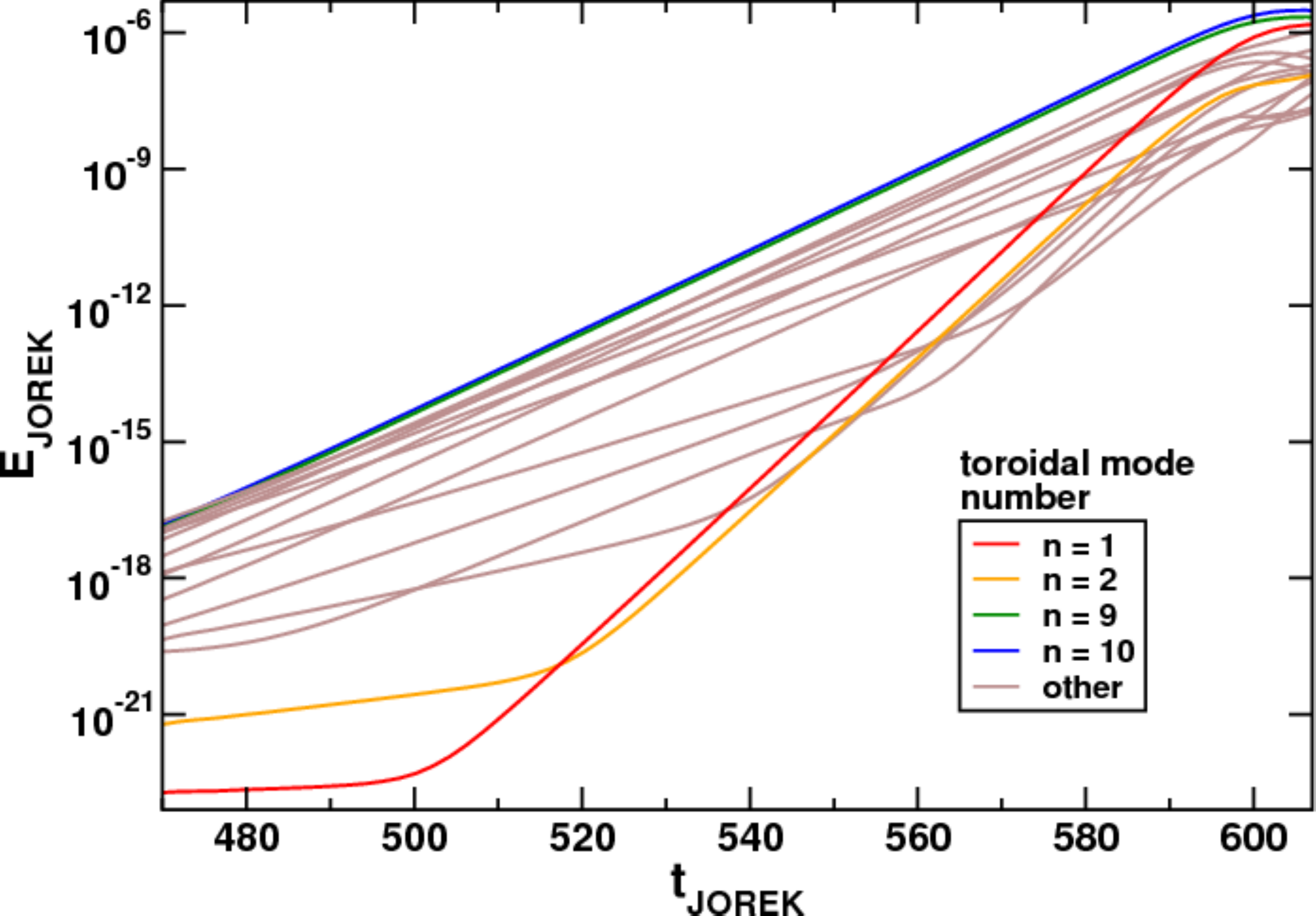}
\caption{Time evolution of the energies contained in the different toroidal Fourier harmonics in the early ELM phase of a simulation with included mode numbers $n=1,2,...,16$. The linearly dominant harmonics are $n=9$ and $n=10$. Energy is transferred from the dominant to the linearly subdominant harmonics, like $n=1$ or $n=2$, by nonlinear interaction.}
\label{fig:energy}
\end{figure}

\begin{figure}
\centering
  \includegraphics[width=0.6\textwidth]{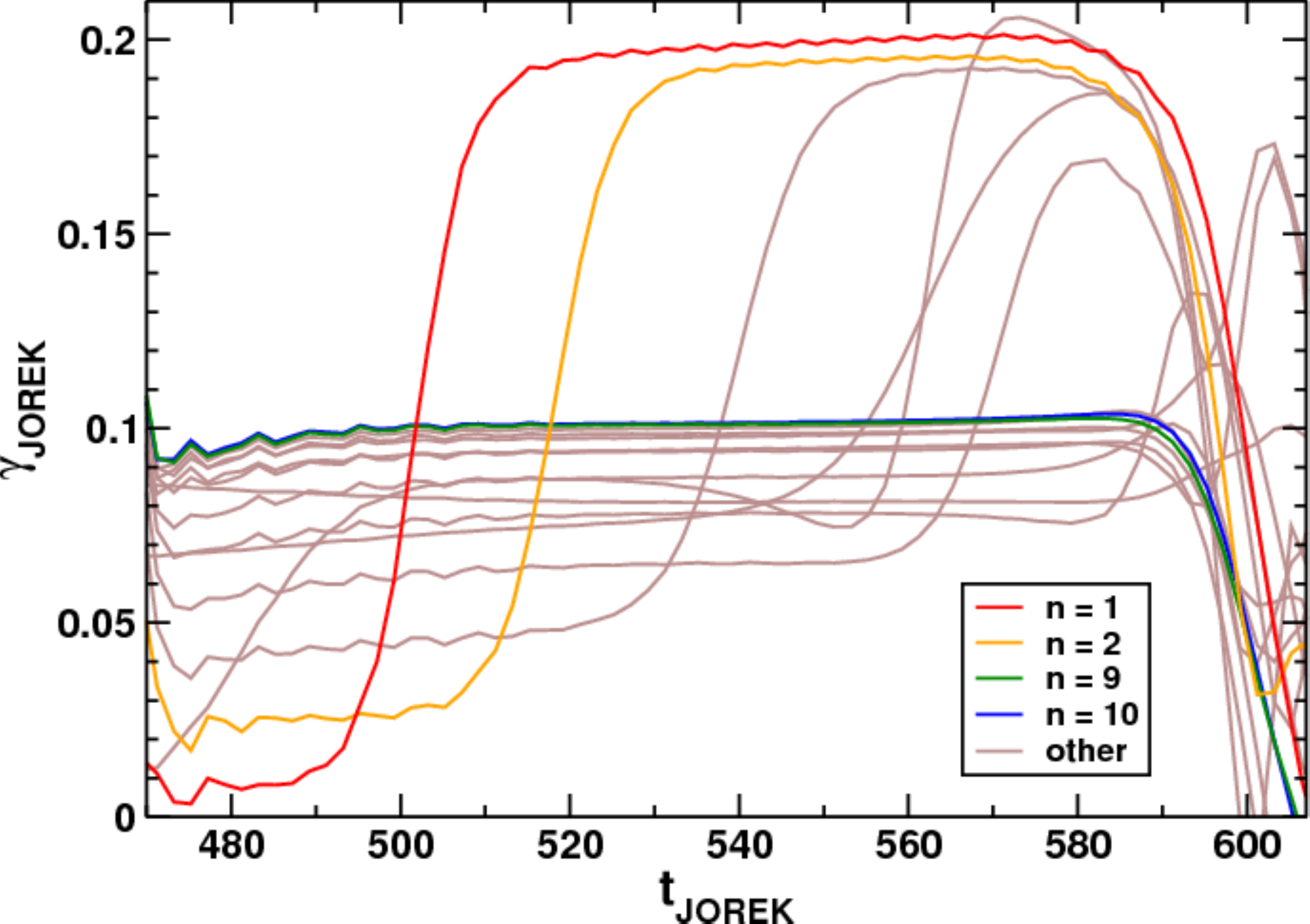}
\caption{Growth rates of the toroidal Fourier harmonics in a simulation with $n=1,2,...,16$. In the linear phase at the beginning of the exponential growth of the perturbation, the harmonics grow at constant growth rates and independently of each other. Subsequently, the growth rates of the linearly subdominant harmonics increase due to nonlinear interaction between the different toroidal harmonics. At the end of this early nonlinear phase, the growth begins to saturate.}
\label{fig:growth}
\end{figure}

The time evolution of the toroidal Fourier harmonics of the perturbation in the early phase of an ELM can be subdivided into three phases, a linear phase, an early nonlinear phase and the nonlinear saturation. Time traces of total energies and growth rates of the different toroidal harmonics in a simulation with 16 included harmonics ($n=1,2,...,16$) are shown in Figures \ref{fig:energy} and \ref{fig:growth}. 

At the beginning of the exponential growth of the instability, the toroidal harmonics grow at a constant rate. The growth rate of a toroidal Fourier component in this linear phase of the evolution is the same as in a simulation where this component is the only included one. It is observed that in the linear phase of this simulation, the Fourier components with mode numbers $n=9$ and $n=10$ grow the fastest. In our simulations, diamagnetic drift effects are neglected which would act stabilizing on high-n harmonics. However, the poloidal resolution limited by computational restrictions also reduces the growth rates of harmonics with high mode numbers. Linearly dominant mode numbers in the intermediate range, as we observe them here, are thus in line with linear theory again.

In contrast to the linear phase where the toroidal Fourier harmonics grow independently of each other, the harmonics start to interact in the subsequent early nonlinear phase. Due to nonlinear interactions, energy is transferred among the toroidal Fourier components which influences their growth rates. Following this phase, the nonlinear saturation yields a decrease of the growth rates. The main saturation effect is, that the background current density and pressure gradient at the edge are reduced by the perturbation, which weakens the drive of the instability. Additionally, the stabilizing influence of the ideal wall boundary conditions becomes more important if the displacement of the plasma due to the perturbation becomes significant compared to the distance between separatrix and ideal wall.

In the following, emphasis is put on the dynamics of the early nonlinear phase before the onset of saturation. It is observed that during this phase, growth rates of toroidal Fourier components which are linearly subdominant increase and that in particular the $n=1$ toroidal harmonic even reaches energies comparable to those of the linearly dominant components, which has already been pointed out in \citep{Hoelzl2012}. This relates to very recent experimental observations. During type-I ELMy discharges in TCV (Tokamak \`a configuration variable) the toroidal mode structure of the magnetic perturbations has been found to be often dominated by low mode numbers, in particular by the $n=1$ component \citep{Wenninger2013}. The magnetic diagnostics in ASDEX Upgrade are not suitable for the detection of low-n harmonics\footnote{The pick-up coils in ASDEX Upgrade cover only a part of the toroidal circumference. Full coverage would be required to resolve an n=1 component, as the growth rate of the mode is comparable to the rotation frequency. Moreover, the pick-up coils measure the time derivative of the magnetic field perturbation which reduces the contribution of low-n harmonics to the signal.} such that it is unclear at present if this phenomenon is also found here. 

\section{Simple quadratic coupling model}\label{:model}
The detailed dynamics of the early nonlinear phase, i.e., why the growth rates of the initially subdominant toroidal harmonics increase, at which point in time the rise occurs and how large the growth rates become, can be explained in the framework of "three wave interaction". Considering a superposition of two toroidal harmonics with mode numbers $i$ and $j$, a second order nonlinear term generates harmonics with mode numbers $|i \pm j|$. Hence, energy can be transferred to other harmonics by quadratic coupling. Based on this idea, a simple model describing the time evolution of the amplitudes\footnote{$A_{i}$ is defined as $\sqrt{E_{i}}$, where $E_i$ is the total energy contained in the $i$\textsuperscript{th} toroidal harmonic.} $A_i$ of the $i$\textsuperscript{th} toroidal harmonics can be set up by

\begin{align}
\pderiv{A_i}{t} = \gamma_{i} A_i + \sum^{16}_{j=1} \sum^{16}_{k=1} \gamma^{i}_{jk} A_j A_k \delta(i \pm j \pm k) \hspace{1cm} \textnormal{for } i=1,2,...,16
\label{eq:model}
\end{align}
where $\gamma_i$ are the constant linear growth rates and $\gamma^{i}_{jk}$ are the coupling constants. As the latter describe the spatial overlap of harmonics $j$ and $k$ in the poloidal plane, they can be set constant assuming that the toroidal harmonics grow rigidly without changing their spatial structure. This is indeed the case for all linearly dominant harmonics. The set of coupled nonlinear differential equations \eqref{eq:model} is able to reproduce to a large extent the time evolution of the toroidal Fourier spectrum of the perturbation in the early nonlinear phase of the JOREK simulations. To achieve this, the appropriate linear growth rates and coupling constants have to be chosen.

The linear growth rates $\gamma_i$ can be extracted directly from the linear phase of the JOREK simulations. From simulations with only few included toroidal harmonics, e.g., two linearly dominant ones which nonlinearly drive a third harmonic, the relevant coupling constants can be isolated. Six coupling constants remain, namely $\gamma^{1}_{9,10}$, $\gamma^{2}_{8,10}$, $\gamma^{3}_{7,10}$, $\gamma^{4}_{6,10}$, $\gamma^{15}_{7,8}$ and $\gamma^{16}_{7,9}$.

Whereas the linear terms of Equations \eqref{eq:model} cause an influx of energy into the system (from the axisymmetric $n=0$ part), the nonlinear terms only yield an exchange of energy among the toroidal harmonics and should thus conserve the total energy. If this conservation of energy is taken into account, for each non-zero $\gamma^{i}_{jk}$ also $\gamma^{j}_{ik}$ and $\gamma^{k}_{ij}$ have to be included into the model and
 
\begin{align}
\pderiv{E_{\textnormal{tot}}}{t} = \pderiv{}{t} \sum_{i} A^{2}_{i} \overset{!}{=} 0
\label{eq:energy}
\end{align}
has to be fulfilled at any time by the system of equations \eqref{eq:model} omitting the linear terms. Equation \eqref{eq:energy} results in additional constraints for the coupling constants such that, taking into account energy conservation, twelve free coupling constants remain. As will be seen later, the additional terms necessary to ensure energy conservation only play a role at the very end of the early nonlinear phase.

The free coupling constants can be obtained by fitting the time evolution of the energies contained in the toroidal harmonics described by Equations \eqref{eq:model} to those resulting from a JOREK simulation. Initial values for the coupling constants are taken from the simulations with only two or three included toroidal harmonics. In every step of the fitting procedure, the system of nonlinear coupled differential equations is solved and the quadratic differences of the logarithmic energies for every harmonic and for a large set of points in time are summed and minimized.

\begin{figure}
\centering
  \includegraphics[width=0.6\textwidth]{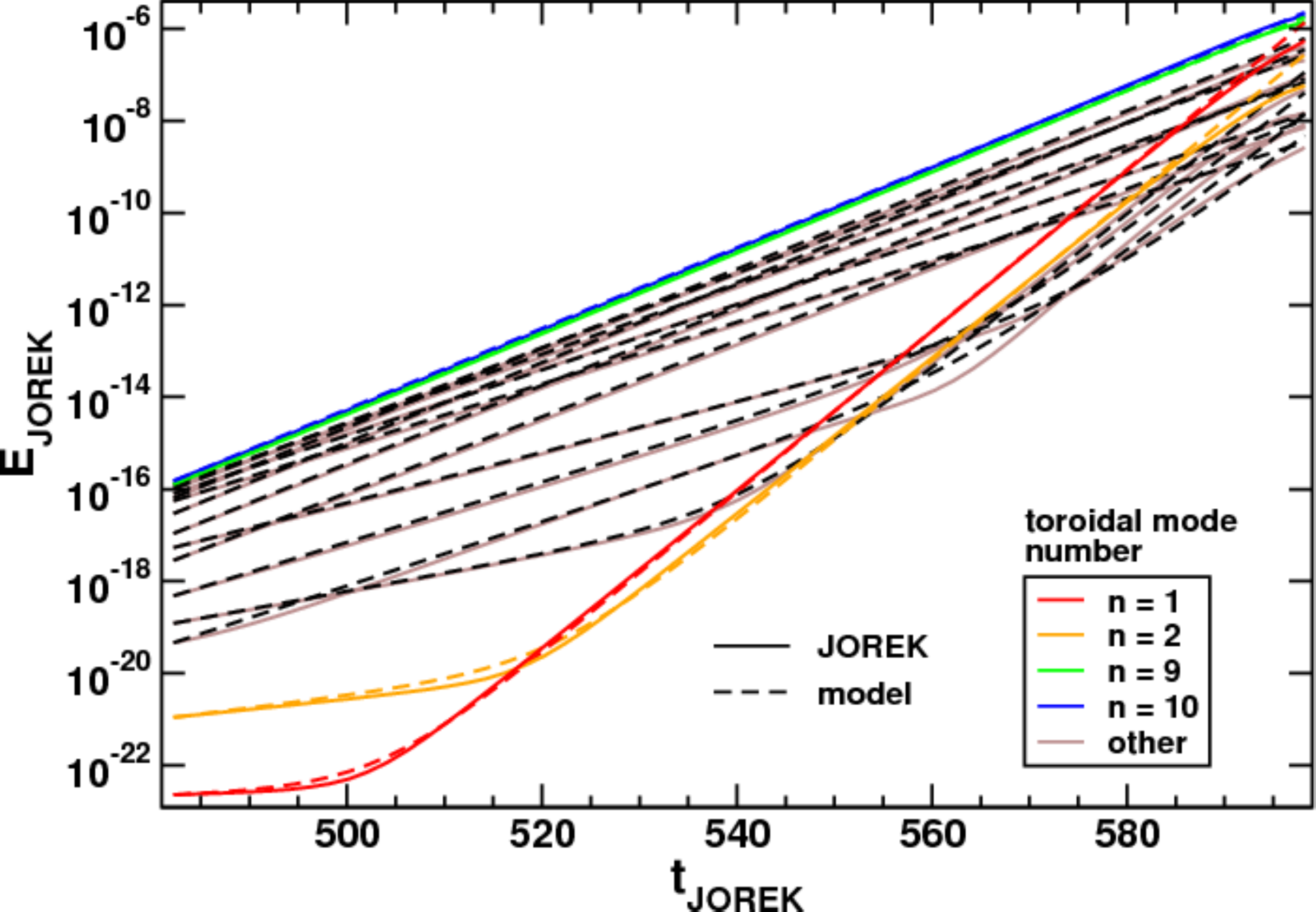}
\caption{Time evolution of the energies contained in the toroidal Fourier harmonics of a JOREK simulation with included mode numbers $n=1,2,...,16$ (straight lines) compared to the results of the simple model with six free parameters (dashed lines). The model is based on the idea that energy is transferred among the toroidal harmonics due to second order nonlinear interaction between them. In the early nonlinear phase, the results from the simple model agree very well with the JOREK results. The deviations between JOREK simulation and model at the end of the early nonlinear phase correspond to the expectations as in this phase the growth is already influenced by saturation effects which are not described by the simple model.}
\label{fig:energymodel}
\end{figure}

\begin{figure}
\centering
  \includegraphics[width=0.6\textwidth]{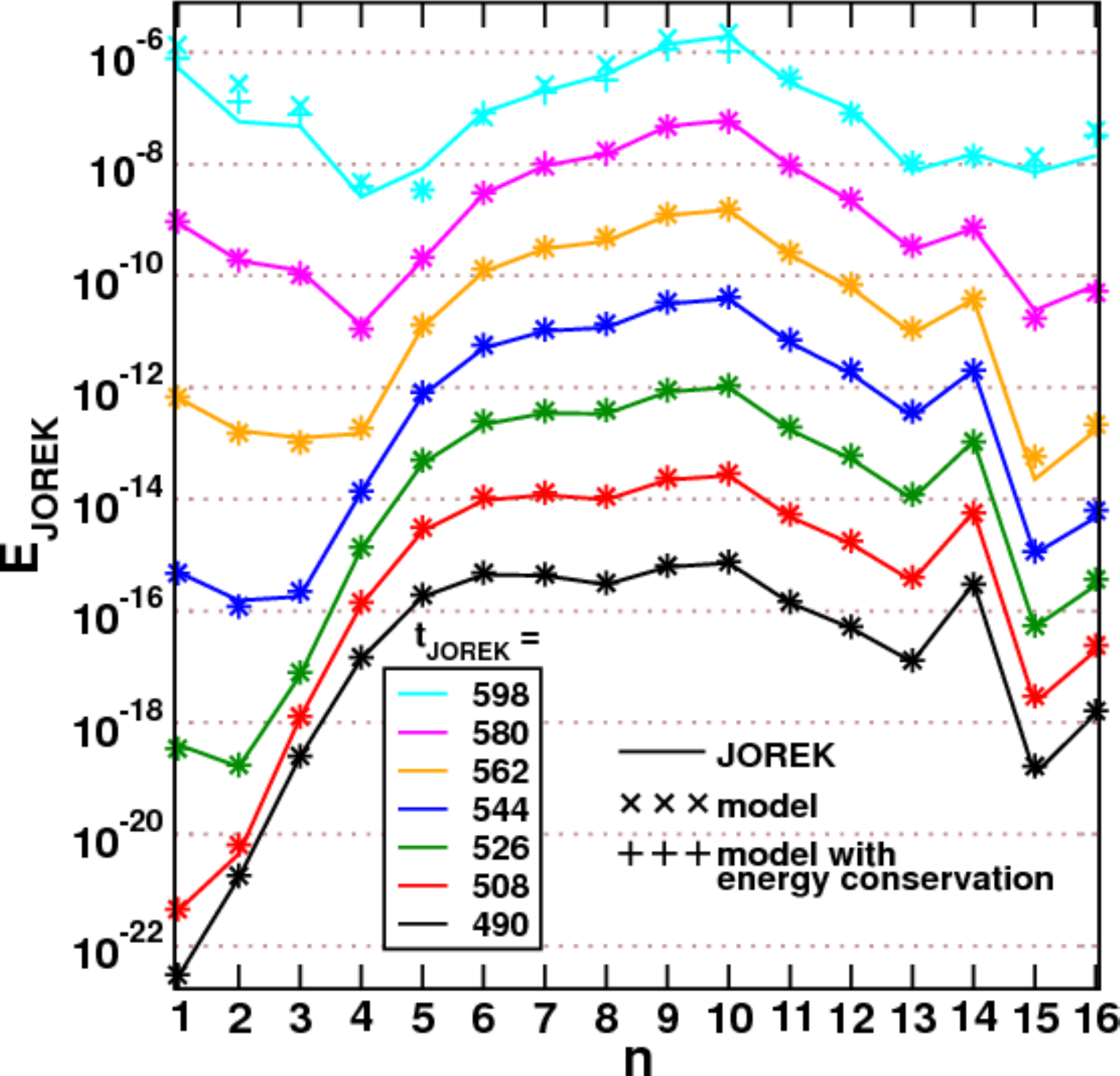}
\caption{Energy spectrum of the toroidal Fourier harmonics in a JOREK simulation with $n=1,2,...,16$ (straight lines) compared to the results of the simple quadratic coupling model (crosses) for different points in time. It can be seen how the low-n part of the spectrum increases significantly and the energies become comparable to those of the linearly dominant harmonics. The plot shows the results of two different versions of the model, one has six free coupling constants (x) and the other one has twelve free coupling constants in order to account for energy conservation (+). The results of the simple model do not deviate from the more accurate one \changes{except} at the end of the early nonlinear phase.}
\label{fig:spectrummodel}
\end{figure}

Figure \ref{fig:energymodel} compares the energy time traces of the JOREK simulations to the simple model with six free parameters. It can be seen that the results of the simulation in the early nonlinear phase are very well reproduced by the simple quadratic interaction model. The values for the six coupling constants obtained from the fit\footnote{The coupling constants obtained from the fit are $\gamma^{1}_{9,10}=113$, $\gamma^{2}_{8,10}=76$, $\gamma^{3}_{7,10}=65$, $\gamma^{4}_{6,10}=21$, $\gamma^{15}_{7,8}=32$ and $\gamma^{16}_{7,9}=34$ (units omitted).} are close to the initial values verifying that the relevant coupling constants were taken into account. \changes{From the excellent agreement using only few free parameters, it can be concluded that the early nonlinear evolution of the toroidal Fourier spectrum is indeed determined by quadratic coupling. Furthermore, it can be seen that the nonlinear growth of a driven harmonic is mainly dominated by one single nonlinear coupling term only\footnote{\changes{This corresponds to the expectations as among competing exponentially growing terms, the one with the highest growth rate will always dominate after a short period of time.}}.} As a linearly growing harmonic evolves as {$A_{j}(t)=A_{j,0} \exp(\gamma_{j}t)$}, it follows thus from Equation \eqref{eq:model} for the growth rate of the nonlinearly driven harmonic that $\gamma_{i,\textnormal{nonlinear}}=d \log A_{i}/dt=\gamma_{j}+\gamma_{k}$, i.e., the nonlinear growth rate of the driven harmonic equals the sum of the growth rates of the two driving harmonics.

The time evolution of the energy spectrum from JOREK (solid lines) and from the simple interaction model (x) are shown in Figure \ref{fig:spectrummodel}. It is visible that at the end of the early nonlinear phase where the saturation sets in, the results from the simple model start to deviate from the JOREK results. This corresponds to the expectations, as the mechanisms responsible for the saturation are not described by the model. \changes{As the linear growth rates are assumed to be constant in the model, a reduction of the drive of the instability due to the effects described above cannot be reflected. Moreover, the assumption of rigidly growing harmonics leading to constant coupling constants, breaks down when saturation sets in.} Figure \ref{fig:spectrummodel} also shows the results from the simple interaction model accounting for energy conservation (+). It can be seen that the additional terms only play a role at the end of the early nonlinear phase. 

The simple interaction model has also been tested on two JOREK simulations with only four included toroidal harmonics ($n=4,8,12,16$) and different distances between plasma and ideal wall which effectively changes the linear growth rates, but preserves the spatial structure of the toroidal harmonics. The results of both JOREK simulations are reproduced well with the simple model by only adapting the linear growth rates but keeping the same coupling constants.


The very good agreement indicates that the simple model provides a good explanation of the non-linear drive of low-n harmonics in the JOREK simulations and could well explain the observations of strong low-n harmonics in the experiment \citep{Wenninger2013}. 

\section{Evolution of the n=1 spatial structure}\label{:n1}
In the previous section it has been shown how energy is transferred to linearly subdominant toroidal Fourier harmonics via nonlinear coupling of the dominant harmonics. It has been shown that the $n=1$ toroidal component can even become one of the dominant harmonics, driven by the interaction between the linearly most unstable toroidal harmonics ($n=9$ and $n=10$ in this case). In this section, the question is addressed, how the spatial structure of the $n=1$ harmonic in the poloidal plane is affected by this energy transfer.

\begin{figure}
\centering
  \includegraphics[width=1.0\textwidth]{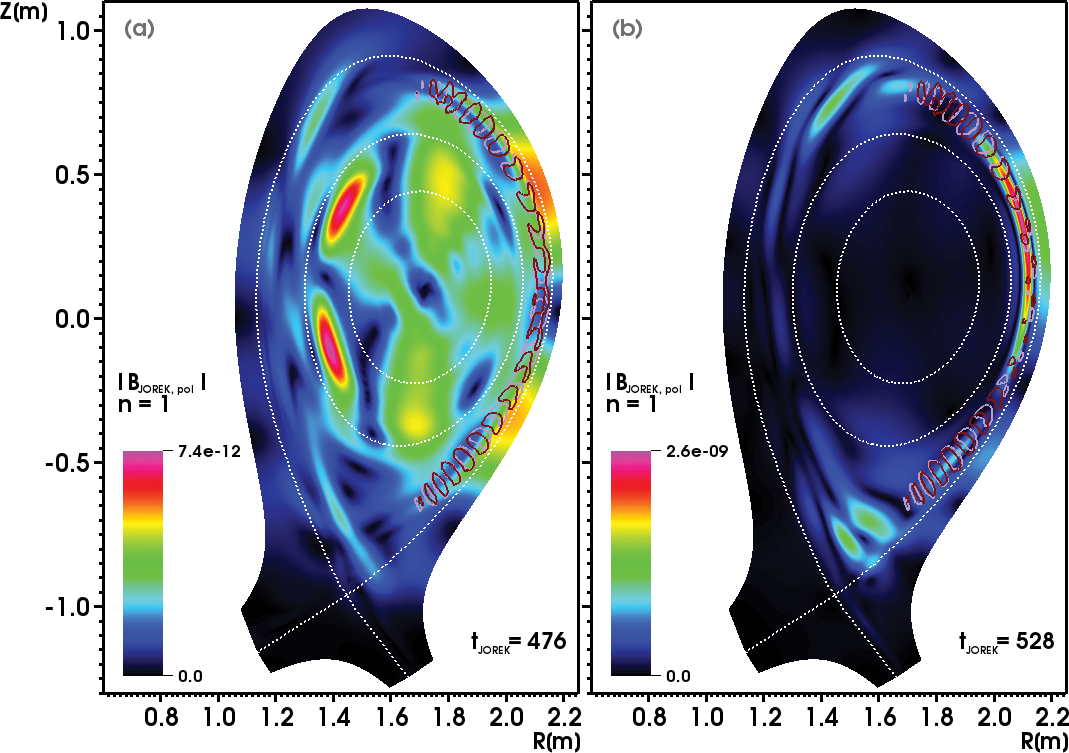}
\caption{Poloidal cross section of the absolute value of the $n=1$ poloidal magnetic field perturbation in the linear phase (a) and in the early nonlinear phase (b) of a simulation with included mode numbers $n=1,2,...,16$. The dotted white lines show the separatrix and flux surfaces at $\Psi_{N}=0.33$ and $\Psi_{N}=0.66$ where $\Psi_{N}=(\Psi - \Psi_{\textnormal{axis}})/(\Psi_{\textnormal{separatrix}}- \Psi_{\textnormal{axis}})$ is the normalized equilibrium poloidal magnetic flux. Contours at 50\% of the maximal value of the absolute value of the poloidal magnetic field perturbation are plotted in mauve for the $n=9$ component and in dark red for the $n=10$ component for comparison. In the linear phase, the $n=1$ toroidal harmonic extends far into the plasma core. In contrast, in the early nonlinear phase, it is radially localized at the plasma edge where also the $n=9$ and $n=10$ are maximal. The poloidal position of the $n=1$ on the low-field side in this phase corresponds to the poloidal region where the $n=9$ and $n=10$ are in phase.}
\label{fig:linnonlin}
\end{figure}

Figure \ref{fig:linnonlin} (a) shows the absolute value of the $n=1$ component of the poloidal magnetic field perturbation in the linear phase of a JOREK simulation with $n=1,2,...,16$. In the linear phase, the $n=1$ component extends over a large part of the whole plasma volume. In simulations where the $n=1$ harmonic is the only included toroidal harmonic, the perturbation grows rigidly preserving this structure\footnote{Note, that at very low energies at the beginning of the growth, the structure of the $n=1$ is still oscillating both in the simulation with $n=1,2,...,16$ and in the simulation with only $n=1$. But the further time evolution of the $n=1$ only simulation shows, that the structure plotted in Figure \ref{fig:linnonlin} (a) later becomes a rigidly growing structure.}. 

\begin{figure}
\centering
  \includegraphics[width=0.5\textwidth]{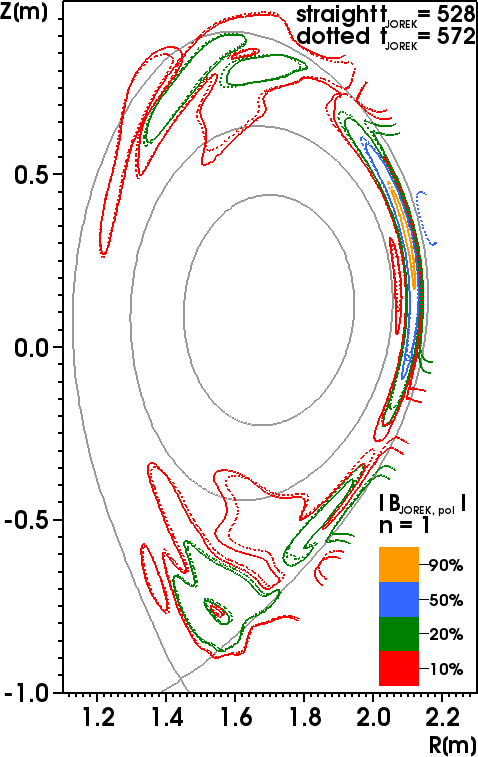}
\caption{Contours at different fractions of the maximal absolute value of the $n=1$ component of the poloidal magnetic field perturbation. The structures at the beginning of the early nonlinear phase (straight lines) and shortly before nonlinear saturation sets in (dotted lines) agree very well, which shows that the structure of the $n=1$ toroidal harmonic shown in Figure \ref{fig:linnonlin} (b) grows rigidly until the onset of saturation. The grey lines show the separatrix and flux surfaces at $\Psi_{N}=0.33$ and $\Psi_{N}=0.66$.}
\label{fig:contournonlinear}
\end{figure}

In contrast to the simulations with only one included toroidal harmonic, in the simulations with $n=1,2,..,16$ the structure of the $n=1$ does not continue to grow rigidly. When the growth rate of the $n=1$ starts to increase due to nonlinear coupling, its structure changes significantly. After a phase of transition a new rigidly growing structure is observed. The rigid growth sets in when the growth rate of the $n=1$ is fully determined by the energy transfer from dominant harmonics. This new $n=1$ structure in the early nonlinear phase is shown in Figure \ref{fig:linnonlin} (b). It is now peaked at the edge of the plasma, in the radial region where also the $n=9$ and $n=10$ are localized. The poloidal localization of the $n=1$ on the low-field side coincides approximately with the region where the $n=9$ and $n=10$ are in phase\footnote{The poloidal angle where this is the case of course depends on the chosen toroidal position.}. The rigidity of the new $n=1$ structure is illustrated in Figure \ref{fig:contournonlinear}, where contours of the absolute value of the $n=1$ poloidal magnetic field perturbation are drawn for two different points in time during the early nonlinear phase. 

The observed evolution of the $n=1$ spatial structure can be interpreted as a superposition of two rigidly growing structures. The first one, visible in the linear phase of the simulation, is the linearly unstable $n=1$ growing at a very small growth rate. The second structure, which emerges in the early nonlinear phase, corresponds to a different, linearly stable but nonlinearly driven $n=1$ which quickly covers the linear structure due to the much stronger growth rate. The phase of transition can indeed be approximately reproduced by superposing the two rigid structures starting at different initial amplitudes and growing at different growth rates.

\section{Conclusions and Outlook}\label{:summary}
Nonlinear reduced MHD simulations of the early ELM phase based on ASDEX Upgrade parameters have been presented. In order to analyze the evolution of the toroidal Fourier harmonics, emphasis has been put on simulations including a large set of toroidal harmonics and simulations including different combinations of these harmonics. It has been observed that linearly weakly unstable toroidal harmonics can achieve large growth rates due to nonlinear coupling of dominant harmonics. In particular the energy of the $n=1$ harmonic becomes comparable to energies of linearly dominant harmonics in the course of the nonlinear phase, which corresponds to recent experimental observations in TCV \citep{Wenninger2013}. To explain what determines this nonlinear behavior of the toroidal Fourier spectrum, a simple quadratic interaction model has been set up, based on the idea that second order nonlinear coupling between toroidal harmonics can generate harmonics with "sum and difference mode numbers". This model is able to reproduce to a large extent the time evolution of the toroidal energy spectrum in the early nonlinear phase of JOREK simulations before saturation sets in. In particular, the model reproduces the growth rates of the linearly driven toroidal harmonics in the \changes{early} nonlinear phase. \changes{This shows, that the nonlinear evolution of the toroidal Fourier spectrum in this phase is predominantly determined by quadratic coupling. The growth of the $n=1$ harmonic is driven by interaction} between the two linearly most unstable toroidal harmonics. 

Furthermore, it has been investigated how the spatial structure of the $n=1$ in the poloidal plane is modified by the energy transfer to the $n=1$ in the nonlinear phase. It has been observed that the rigidly growing structure of the linearly unstable $n=1$ which reaches far into the plasma core transitions into another rigidly growing structure of a linearly stable but nonlinearly driven $n=1$. This second structure is localized at the edge of the plasma, in the region where also the two linearly dominant harmonics are maximal, which is in line with the idea that the $n=1$ emerging in the nonlinear phase is generated by the interaction between these harmonics. The assumption brought up in Reference \citep{Wenninger2013}, that a strong $n=1$ component gives access to the plasma core, which could explain the large losses of energy observed during type-I ELMs, is thus not supported by the simulations, as nonlinearly, the $n=1$ becomes highly localized at the edge. Nevertheless, strong low-n components could couple easier to core instabilities having a similar toroidal structure, such as neoclassical tearing modes.

As a next step, it would be interesting to render simulations with more realistic values for viscosity and resistivity possible. \changes{As diamagnetic stabilization and sheared toroidal plasma rotation are expected to have some influence on the nonlinear coupling between the toroidal harmonics, including these effects in the simulations is also planned.} Moreover, simulations exceeding the early ELM phase are subject of ongoing work.  

\section{Acknowledgments}\label{:ack}
A part of this work was carried out using the HELIOS supercomputer system at Computational Simulation Centre of International Fusion Energy Research Centre (IFERC-CSC), Aomori, Japan, under the Broader Approach collaboration between Euratom and Japan, implemented by Fusion for Energy and JAEA. This work was partially funded by the Max-Planck/Princeton Center for Plasma Physics. K.L. would like to acknowledge the support by the Austrian Science Fund (FWF) under Grant No. P19901. The authors would also like to thank Ronald Wenninger for the discussions about the experimental findings.


\end{document}